# Evolution of science II: Insights into working of Nature[1]

Mayank N Vahia

Tata Institute of Fundamental Research, Mumbai

vahia@tifr.res.in

Abstract

We attempt to provide a comprehensive model of evolution of science across millennia taking into account the contributions of other intellectual traditions, cultural value system and increasing in sophistication of humans in their study of nature. We also briefly discuss the role of technology and its interplay in the evolution of science. We identify five primary approaches to the study of nature, namely ad hoc formulations, religious approach, pragmatic approach, axiomatic approach and the logic based approach. Each of these approaches have had their prime periods and have contributed significantly to human understanding of nature and have also overlapped within a society. Each approach has had a central role over human evolution at some stage. We surmise that the currently dominant axiomatic method will reach its limits due to complexity of the system and may never be fully formalised. We suggest that the future progress of science will more be a logic based approach where experimentation and simulations rather than axiomatic firmness will be used to test our understanding of nature.

1. Human studies of nature

In Vahia (2015) we had analysed the evolution of human intelligence and perspective on nature that ever since humans – even archaic humans – obtained intelligence beyond their survival needs, they began to investigate nature to improve their living conditions. There is significant evidence of development of technologies by archaic humans and of careful burial of the dead by Neanderthals suggesting a certain world view and respect for the dead. Here we discuss the evolution of the thought process that has been resulted in our present perspective.

Early humans would not have been able to comprehend the variety in nature. Even modern humans cannot make such a claim. However, they must have been able to see that there were rhythms and consistency in the working of nature but that these patterns were not exact. However, given the nourishing nature of land and the need for rains, their first instincts seem to have been to relate Earth to Mother and Sky to Father. But Mother Earth and Father Sky also had patterns which were not perfect and the causes of these deviations were difficult to fathom. It is at this stage that the idea of God would arise. Mothee Earth figurines are amongst the earliest known artwork (Conard, 2009). Many early cultures show rock art with a human form holding Sun and Moon in two hands.

---
[1] Submitted to Current Science, June 2015

2. Role of technology in the evolution of science

Technology and science have been feeding each other to their mutual benefits. In early human manipulation of nature, technology probably preceded analytical studies. The growth of human technological capabilities are discussed elsewhere (Vahia, 2015, figure 1). It shows the relation between technologies of scientific discoveries. A typical scientific discovery gets gradually converted to technology, and these technologies open up various possibilities leading to next set of useful technologies. Civilisations progress by the effective technology that science provides and not by novel scientific insights alone. The early technological evolution, from early stone tools to construction of dwellings, were developed from an instinctive understanding of nature. Even these, especially the skills needed to create flaked tools, animal traps, controlled fire etc. would require a certain basic understanding and acceptance of the objective nature of the environment. While scientists take great pride in the impersonal nature of their work, cultural influences play a significant role (Iaccarino, 2003). While technology plays a crucial role in the evolution of a society (figure 4), it also provides new insights into the working of nature. For example, the realisation of the power of steam to do work eventually led to the field of thermodynamics. There are several such examples in science. In the present discussion we do not discuss this subtle interaction between the two and integrate both, technological and scientific advancement into a single unit.

There is no denying the elegance in the working of nature. To begin with, repeatability of a property of physical universe, conservation of matter and other evidence of natural consistency would have given them faith to investigate nature even further. A section of human intellect was therefore always directed towards identifying patterns and keeping count. While counting can start with commerce and then grow into complex ideas, geometry is essentially a gift of astronomy. This systematic study of quantifying the working of nature would have had a profound effect on humans. Different cultures have approached the study in different manner (see e.g. Narasimha 2003, Ganeri, 2001, Wilder, 1960). These studies had different approaches:

1) Ad hoc Approach
2) Religious Approach
3) The Pragmatic Approach
4) The Axiomatic Approach
5) The Logic-based Approach

We discuss each of them in detail below. In table 1 we give a summary of the different methods. While we have classified these methods for convenience, many of these approaches have overlapped in different cultures and the path is not monotonic. For convenience and in keeping with a more broad approach we have ignored culture specific variations and evolutionary paths. To illustrate the differences, in table 1 we give an example of how the cultures would treat the observations of fire on a hill. With each approach we illustrate the characteristics of the culture with examples from astronomy. The relevant astronomical techniques imply a whole host of other technological developments but, for the sake of brevity, we do not discuss them.

Ad hoc Approach: This purely utilitarian approach is entirely driven by survival needs and instinctive understanding of the properties of material employed to improve survival. No systems are formally studied and little formal planning is included. Systems are built based on intuitive feel and experience of material and their combination to create the necessary tools. All science begins this way. While this may be called primitive, a significant amount of informal understanding of material is required to be efficient. During this period, one typically finds advent of rock art with astronomical theme and megalithic structures designed to keep track of the movement of the sun. The method is still prevalent in many low technology activities.

Religious Approach: This approach assumes that the nature and the universe is driven by a supernatural power who tracks everything and controls all events and the evolution of any event is based on the whims of the supernatural being (see e.g. Culotta, 2009). The wishes of this supernatural being are dependent on the nature of human behaviour and has serious problems on issues such as free will and the manipulation of the living by the superhuman. As such, it discourages any analytical study of nature and encourages expenditure of time and resources to ensure that the superhuman remains positively disposed towards the humans. It therefore aggressively denies and discourages formal studies of the working of nature. It can also give rise to irrational belief systems and occasionally hide analytical studies of nature within its reach by giving it a different perspective. It has also moulded and changed the manner of growth of civilisations. The extent to which dominance of scientific method can be negated by religion can be seen today in many West Asian countries, which began its history by encouraging scientific thought but finds its scientific approach completely stifled by the rise of religious dominance severely restricting its future prospects (Hoodbhoy, 1991).

During this period, the most prominent feature is the evolution of megaliths into sites of important religious or semi religious festivities, chiselled rock art as well as rise of myths connecting heroes, gods and heavens. Elaborate stories of the times when gods ruled the earth and interacted with humans are created and rituals are designed to keep the gods happy.

However, the interplay between religion and science has often been complex since religion also evolves with time (Wade, 2015) and many scientists would pay their respects to the elegance of science. The rationalist approach to life and universe is often not easy to escape and atheism is often not easily accepted largely driven by the manner in which human intelligence has evolved (Boyer, 2008). Many scientists such as Isaac Newton were involved in religious studies or have been practicing formal religions. Ball (2008) has discussed at length the relation between science and religion and the manner in the mutually differing emphasis on the core entities that govern the world have been handled by human civilisation.

The Pragmatic Approach: This approach assumes that nature works with mathematical precision but its exact nature of why she does so is beyond complete comprehension. It implicitly assumes that complete comprehension about why nature behaves the way it does is beyond comprehension. With increasing levels of comprehension, more subtle variations appear. As such, any mathematical formulation was an approximation of nature, valid till a better approximation – that fitted the observations better – is found. All knowledge is ad hoc and transient representation of nature. Almost all cultures began their study of nature implicitly or explicitly, with this premise and most continue with this premise. This approach allowed them to take up everything from complex architecture to accurate positional astronomy. Note that this approach also relied extensively on mathematical representation but assumed to be an approximation. The biggest advantage of the pragmatic approach is that it provided a way around the suffocating hold of the religious approach to science and avoided the direct conflict with religious ideas that has marked the axiomatic approach to the study of nature.

During this period, a sense of autonomy amongst the learned results in elaborate observations of nature and mathematical modelling of the working of the universe. Epicyclical movement of planets and the corresponding geometrical and algebraic ideas as well as measurements of the size of the earth etc. are typical exercises that are taken up during the period.

The Axiomatic Approach: This approach assumes that nature works in strictly logical way. It is therefore possible to understand nature by separating different aspects of the working of nature and studying them in isolated environment. The Greeks were probably the first to be obsessed with this idea and became committed to these ideals. However, in the absence of good data – or even good pragmatic ideas – their axiomatic approach did not progress beyond the works of Archimedes and other Greek scientists. It remained alive only as a noting of interesting ideas in the forgotten or lost Greek tests and Arabic culture and did not find much favour in India. During this period, the entire set of ideas on how the universe has been seen to be working are formalised and a demand for logical consistency is made on the working of nature. During the period, astronomers formulated ideas of gravity whose formulation depended on early observational records of the pragmatic period. They then merged it with the realisation of conic sections as the shapes of orbits and provided the first physical model of the solar system and gave glimpses of the universe beyond. Developments in physics and other fields opened the doors for multi wavelength and telescopic observations of the universe. Typical theoretical study would involve idealised, simplified analysis of real physical systems, often simplified to fit into the mathematical capabilities of the period.

The Logic-based Approach: This approach assumes the working of nature had certain underlying principles which are subject to analysis, but isolated mathematically formulated principles only have limited applicability. In reality nature is complex and not amenable to the classical axiomatic formalism. So while one can still create mathematical models of the working of a small aspect of nature they will not be central to understanding to nature. The underlying physical ideas will be provided by specific assumptions valid for the particular problem being addressed. In many cases, the linguistic format may be more conducive format for understanding nature. By implication therefore the set of axioms and formalisms that explain nature will not be a finite set but will consist of an open ended vocabulary. This language will be precise in its definition of words and the formulation of linguistic structure will have precision of consistency and structure. The words and grammar of the language will be traceable back to a set of rules. The rules of modification for application to a local situation will be logical and intrinsically explanatory as well as subject to rigorous but descriptive or informal logic. This approach will subsume the Pragmatic Approach (and will generalise the Axiomatic Approach) by description that will to have a visual impressionist approach to the behaviour of nature. Saturated by approximate correlation between theory and experimental data, astronomers begin to appreciate that their early simplified analysis that allowed analytical solutions to observations no longer provide the complete description of reality and including more realistic information takes the problem beyond the capabilities of elegant analytical solutions and theoretical studies are either approximated or simulated to provide better insights.

With the advent of formal mathematics this idea of informal logic would expand to formal logic where propositions cannot to be proven to be correct from the initially basic rules or operation. The study of nature will put greater emphasis on geometry, analysis and logic and the classical, algebraic approach would have reduced applicability.

1. Comparison of different approaches

Some basic scientific understanding is evident and common to human development that arose before humans dispersed all over the globe eighty to a hundred thousand years ago (Vahia, 2015). These include cave making, cave painting and possibly some basic ideas of early religion. These are common in many early cultures in different parts of the world. Some form of language probably existed much earlier (Dediu and Levinson, 2013) but the diversity within these languages is significant and suggests that a large fraction of the development of language was local to different regions (Evans and Levinson, 2009).

However, early approaches were a mix of ad hoc and pragmatic approach. For the purpose of our analysis of evolution we will not delve on the ad hoc approach since it naturally progresses into pragmatic approach with the advent of education. The religious approach is similarly an intellectual dead end, the exploration of ideas, theories, images and myths about this superhuman and his creation have commanded a significant amount of human intellectual resources and continues to do so. However, we shall ignore this line since it does not even attempt an analytical approach to understanding nature.

Most cultures have used this approach in understanding the working of nature. Interested reader is referred to Chagett (1995) for Egyptian science and Subbarayappa (2007) for Indian science. The most exhaustive of these studies is a series of volumes by Needham (1954) has discussed Chinese science at length. All these studies suggest that the pragmatic approach was adopted according to continuing advancement in mathematical astronomy driven by cultures and people not particularly sensitive to religious approach. However, their focus remained on identifying and applying new and needed technologies for the general wellbeing. A specific idea of classifying the working of nature does not seem to have been the focus of these studies. The classical approach of this kind of studies was to classify nature into four or 5 basic entities namely solid liquid, gas, energy and sky. Amongst the most detailed approach is the one explored by the Indian civilisation around 600 BC. This included classifying nature not only into 5 basic properties and assigning various attributes to the same (Figure 2) that appears in the book Vaiseshika of Kanada (Mishra 2006; Chakrabarty, 2003; Chattopsdhyaya, 1912). For a more general discussion on Indian philosophy and philosophy of science see Sarukai (2008).

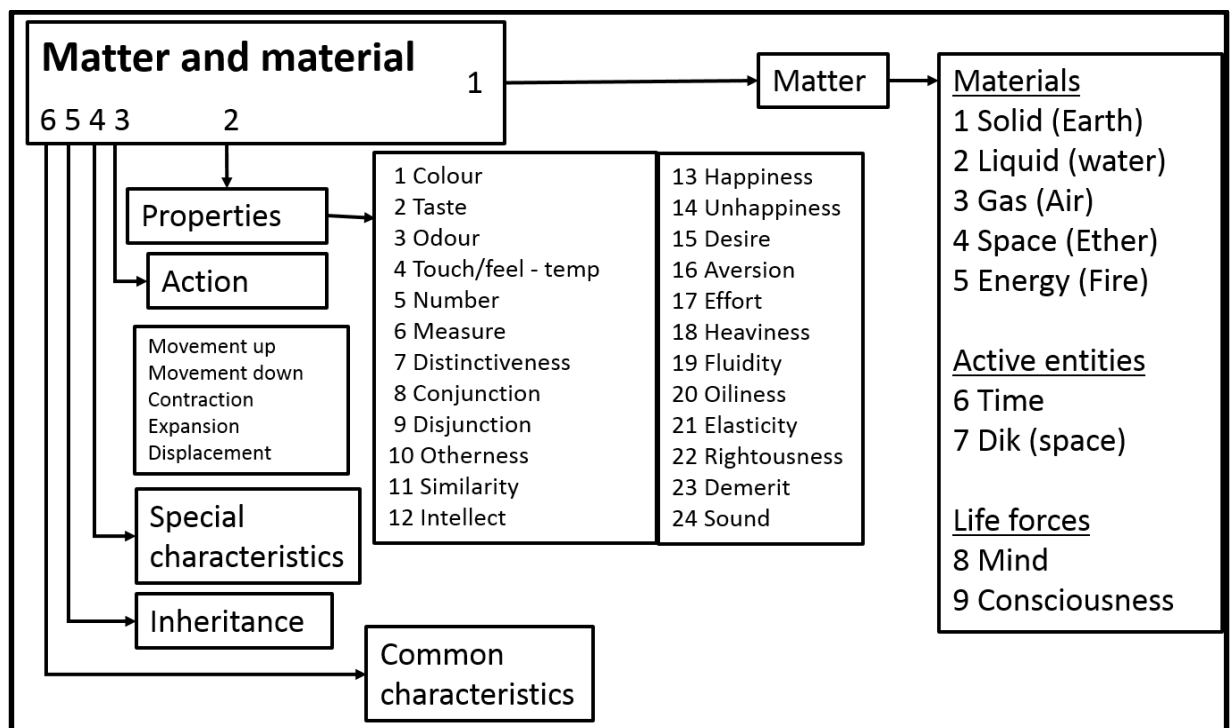

*Figure 1: Organisation of nature in Indian philosophy*

The sophisticated enough to explain various properties of matter and their changing form including mechanics etc. and did not need intervention of unknown forces to intervene in the working of nature. However, it was not extended to search for underlying physical laws that governed the universe. No attempt was made to understand the underlying principles routine situations or in the application of a technology. Hence fields like classical mechanics (which is a purely analytical study of mechanical properties of matter in isolated systems) and thermodynamics (which arises from study of gases) were never pursued and would probably not have been pursued at all. This is interesting because mechanics and chemistry were fairly advanced and Chinese even harnessed steam energy. So they built everything from the Great Pyramids and Taj Mahal and steam engines but did not worry about the roots of basic properties of nature. This narrow focus put rather stringent limits on how far this field would have progressed.

The axiomatic approach has been the most perceptive of all these approaches and in common perception it is often assumed to be the beginning of the scientific revolution and beginning of the scientific approach to life. However, it is worth recalling that the axiomatic method would not have worked in isolation. It needed long traditions of meticulous observations that predated the Renaissance period when this method flourished. Without a massive amount of universal understanding of the nature the axiomatic method would have failed – as it did in early Greek period. Its primary success was in applying it to all aspects of the working of nature. Its most spectacular success has not been so much in technological innovation as in the realisation that the earlier approaches had been ignorant of some major aspects of human studies. Starting with Galileo's astronomy to Newtonian mechanics, it led to the field of thermodynamics and recognition of electromagnetic fields and particle physics along with several allied fields. At the same time, it was responsible for phenomenal increase in technologies. However, as we shall see below, we seem to have reached a plateau in these studies and this approach seems to be self-limiting.

The difference between the logic-based and pragmatic approach is that the logic based approach assumes that mathematical precision is consistent *within an underlying, objective, physical framework*. While pragmatic approach demands simple predictability of events based on formulations, logic-based approach insists on underlying logical consistency which may or may not be amenable to formal mathematical approach. This in turn would permit analysis of complex situations where several of the axioms were at play simultaneously. It differs from Axiomatic Approach in that it does not demand a physical basis for the validity of a formulation to be consistent and accurate in terms of defining an environment.

The pragmatic approach was highly successful in its understanding mechanics. However, adaptation of mechanics by the axiomatists required concepts such as friction, centripetal force etc. to meet the demands of axiomatic consistency. For example, friction itself arises from electrostatic forces and any axiomatic study of friction must begin with intermolecular forces. However, the most common approach to using friction is to assume a (measured) ad hoc parameter called the coefficient of friction. Such broad and working generalisations pervade all aspects of science and most physics does not begin with atomic structure but with the idea of 'bulk matter' – which is a pragmatist's approximation lacking the purity demanded by an axiomatist.

Even then, civilisations that were pragmatic in their approach, also worked on mathematical formulation where possible since axiomatic approach has the elegance of simplicity. While studying mathematics, they found that purely formal approach worked well. For astronomy the logic of consistency retained their validity over long periods of time. For example, without gravity or need for Heliocentric or Geocentric models, using mathematical formulation permitted the Indian pragmatists to extend their studies significantly. Using the concept or logic of *prakruti swabhav* (compulsion based on one's nature) for each planet's controlling equation was satisfactory. Indians were so committed to the pragmatic ideas that even while invoking the ideas of epicycles in planetary motion, they used the mathematical formulations without worrying about underlying axioms the way the Europeans did. So while the Europeans were trying to define circles within circles and fitting their radius and trying to explain why this happened, the Indians were quite satisfied with the mathematical formulation and the relative locations of planets where retrograde motion needed to be included. They did not significantly extend their studies to more classical systems and these were left to more ad hoc experimentation.

The absence of search for axioms and satisfaction with pragmatism meant that the description of nature such as astronomy reached a gradual progression and reached a plateau soon. Mechanics was left to the technological development and left isolated from the developments in mathematics. Hence these fields while making solid progress using the concepts from mathematics, did not attract the intellectual investigations on the reasons why these mathematical models worked and must have remained a logic-based delight. They built large and complex architecture and technologies which would have needed an understanding of the underlying mathematics but that did not lead to the trying to figure out what went at the core of *prakruti swabhav*.

Hoodboy (1991) has discussed the issues related to religion and science (in the context of Islam), the pragmatic Muslim approach to science and the western axiomatic approach to science. He points out that the fundamental nature of axiomatic science is its very secular nature in the sense that it deals with worldly matters and accepts no authority. He points out that even at the peak of its success in the Arab world till 13$^{th}$ century before was overwhelmed by orthodoxy, the subject remained elitist following reasons (Hoodbhoy, 1991, p 93-94) in the context of Islamic or Muslim science:
1) The applications of ad hoc science were limited and hence did not enthuse the artisans and tradesmen at large.
2) Since it progressed by court patronage, the focus of the practitioners of science was to please the court rather than design new devices.
3) It never found its way into the teaching curriculum at large and was restricted to a few schools.
4) The authors of great works went out of their way to restrict the readability of their writing so that the commoners did not get to comment on it or access it.

This is probably true of all cultures that practiced pragmatic science. Hoodbhoy (1991, p 118 to 133) also discusses the specific social structure of the Arabs and Muslims who had taken the studies further from the Indian culture did not take it to the next level of axiomatic approach in which the Europeans excelled.

The axiomatic method would not have worked in isolation. Without a massive amount of universal understanding of the nature developed by the pragmatists, the axiomatic method would have failed – as it did in early Greek period. One example of this is as follows. Matter has mass and hence is subject to gravitational pull. Hence humans stand on earth due to gravity. A corollary of this would be that insects crawl on humans also due to gravity. However, this is clearly not true – for insects to be on humans, you need electrostatic forces.

The Europeans in the Renaissance period absorbed the results of the pragmatic approach to mechanics and axiomatic approach to mathematics also learnt of the ancient axiomatic traditions of Greeks (acquired through Arab records) and revived them with vigour

Evolution of science II: Insights into working of Natureeven as they heavily borrowed from logic-based and pragmatic approach of the Asians. Note that purely axiomatic approach of the Greeks had not got very far – it needed crucial inputs from the other methods of study.

The result of these developments was that they had a rich field of data, experience and mathematics that they converted to axiomatic sciences. Since nature responded well to these axioms founded on earlier pragmatic studies, Europe made quick progress in our understanding of nature and our capability to manipulate it. With a commitment of experimentation for validating their axioms they soon discovered thermodynamics and electromagnetism – fields that had been completely missed by the pragmatics – even though they had extensive experience in metallurgy.

However, as we begin to study inherently complex systems where multiple axioms work simultaneously, neither ad hoc localised formalism nor superposed formulation of multiple concepts together will succeed. Science is also looking at system in in real environment which puts additional limitations on development of clear axioms to study nature.

The approach that is now gaining ground is that the working of nature has underlying principles which can be analysed and described in descriptive form. However, these are not general essays but the terminologies used are precisely defined. These precisely defined terms are considered necessary and sufficient to describe some aspect of nature. It is therefore necessary that the words and grammar of the language should refer back to a set of rules. The rules were logical and intrinsically explanatory as well as subject to rigorous if informal logic and, amenable to mathematical approximation. However, mathematics may not be the best way to describe them *in view of the inherent complexity.* Hence it is impossible to prove 'facts' and the best we can do is to state that something seems true based on all available experimental (and simulation data). This approach ran parallel to the Axiomatic Approach and provided analogy for mathematical representation. However, with increasing complexity of problems being addressed, this is now the principle means of understanding nature with simulations stepping into the place of formal proofs. By removing mathematical description, it brought in some much needed approximations in description of nature.

2. Godel, Complexity and the limits of axiomatic approach

There are two primary reasons why the Axiomatic Approach will be self-limited. Detailed studies of science have made it clear that formalising science in the mathematical sense is not easy and may not even be possible (see e.g. Watson, 1963). Axiomatic approach therefore will not be able to encompass the entire set of results in physical sciences in its totality. The natural reality in many cases is inherently complex and driven by fractals and chaotic undercurrents which cannot be fully predefined in an axiomatic manner. Also, bulk studies of matter in particular are further vulnerable to interferences which cannot be modelled from first principle and operative simplicity will need to be employed depending on the scale of the problem and the detailed that need to be or can be understood in a specific situation. However, even if this barrier were to be overcome, Godel's Incompleteness theorem would limit the Axiomatic Approach. We discuss this in detail below.

These are the Godel Wall that arises from the work of Kurt Godel that shows that a purely axiomatic system will have incompleteness problem. Such systems will have to accept facts that it cannot prove. The second limitation arises from the fact that systems are now studied in their full complex and more in *in situ* and realistic environments. These studies it will struggle to prove its validity from first principles and will rely on non-deniability through experimentation and simulation. It will also make it essential to explore fundamentally different ideas about the organisation of nature.

The fact that the Axiomatic Approach will be saturated is clearly demonstrated by the Godel's Incompleteness theorem that states that in any axiomatic system will have statements that even though true, will not be provable within this system (see for example, Nagel and Newman, 1960; Panu, 2014, see Franzen, 2005, 2006 for the limitations on applicability of Godel's theorems to other fields). Godel's work shows that any axiomatic approach is self-limiting. The result is that sooner rather than later, the axiomatic approach will be manifestly incomplete in the sense that they will not be able to prove all statements that are true. However, the axiomatic approach of the study of nature is far more powerful than any earlier approach. But, as systems become increasingly complex, the axiomatic approach will begins reach its limits, and it will no longer be possible to explore nature purely on the basis of axioms since we will begin to encounter systems whose complete description will no longer be provable within the axiomatic system. Future studies will begin to increasingly rely on pragmatic formulations governed by experiments and turn to a more logic-based approach to understanding nature.

This puts a severe limit on the reach of axiomatic science and as long as they claim to represent all aspects of nature. A theory of everything would be a formal system where Gödel's theorem applies, and in such case the system will not be able to provide proof for all that is true even within this system. We will have to accept that in so far as we accept that the basic axioms of science form a total system of a description of the physical world, we will also have to accept that it will not be complete in that it will not be able to prove everything. The alternative is to assume that the axiomatic system is not complete in the sense that there will be systems which it cannot establish from within its set of axioms. In which case, science will never have a complete set of rules and even though its rulebook will be self-consistent (and not internally contradictory), it will not be complete. A system can be consistent but not complete and amenable to analytical studies (Franzen, 2006). Under these conditions the Godel's theorem does not apply. However, these systems then will continuously need additions of axioms to explain the system with increasing complex rules for adaptation, making it unwieldy.

In addition, in the case of physical systems, complexity of such system also does not help the axiomatic approach where too many axiomatic processes work simultaneously. All realistic systems are complex system and not amenable to the kind of simplification crucial to mathematical description. Experimentation, simulation and 'true to the best of our knowledge' approach will dominate.

The result will be a merged field where axioms will not be proven but will be shown to be non-deniable . However, the validity of these descriptions of the governing principle of nature will have to be proven by non-falsifiability within the reach of computer modelling and extensive testing. A concept will be true because it cannot be falsified under any situation we can think of and simulate. Physics is relatively idealised and isolating the systems is possible and hence its growth along axiomatic lines is possible. This is not the case with biology which, at best can use axioms from chemistry but still needs to be validated and the interplay of multiple axioms simultaneously is difficult to judge or generalise. The axiom of chemistry themselves are generalised concepts from atomic physics since it is not possible to revert back to atomicity for every extension of knowledge. It is logician's delight.

The future of science therefore is more and more drifting away from purely axiomatic approach as various subjects reach the Godel Wall. One can argue whether the Godel wall is a limitation of human mind or whether the complex systems (with their intrinsic tendency to be chaotic) are difficult to define axiomatically. String Theory, for example claims legitimacy based more on a logical approach than axiomatic proof. Cosmology is another field where the Godel Wall arising from lack of knowledge of acceptable axioms – has resulted in logical approach to science. The usage of cellular automata and its related modelling (Wolfram, 2011) is one example of this changing emphasis on science where again simulation seems to be the way of validating (or discounting) a scientific hypothesis.

In some sense this is also a reflection of the human brain. Designed to survive in the wild with 3 requirements – to eat, not be eaten and reproduce – human senses are hierarchical with visual sense having the highest priority. This predisposes the brain to visualise and accept a visualised picture as an acceptable expression of the working of nature. Any visualisation eventually become a more logic-based and accepts non falsifiability within the reach of experience as satisfactory proof of validity. So while experiments remain the final arbiters in any rational analysis, the axiomatic approach is easily replaced by logical or even pragmatic approach.

So the future of science is increasingly logic-based and pragmatic. Technologies will work entirely with logic-based technology. The basic argument is that nature obeys a set of rules and they can be combined into a machine which in some ways makes our life more comfortable or interesting. Hence the exponential increase in knowledge may have been triggered by axiomatic approach, it is only a transient state in the long march of humans to understand and master nature for their personal gains.

In the figure 2 below we have attempted to plot the path of growth of science. It is a purely intuitive and off scale plot to aid thinking. It discusses the fields of science that were discovered by various approaches and the broad geographical regions that dominated the fields. It suggests that the Axiomatic Approach is reaching its limit after a strong growth for the past 400 years. We have suggested that the Logical Approach will probably not be as spectacular but this statement is made more based on the experience that any new system of knowledge or approach takes some time before it matures to a level where it can contribute significantly to our understanding of nature and will have to run parallel to the Axiomatic Approach for some time, especially in physics even as the Logical Approach is already visible in other fields of science. Another reason why the Logical Approach may not gather exponential growth is that the hardware and to some extent the software expansion rate is reaching its own limits (Markov, 2014) and unless new generation of ideas such as quantum computers or new approach like the cellular automata (Wolfram, 2001) or such fundamentally different approach arises.

We are now on the threshold of the post axiomatic phase which will require fundamental restructuring of our thinking about nature and science and their mutual complementarity. With no axioms to validate a hypothesis, we will have to redefine how we validate a given experimental result. We will have to have new criteria of reliability of results and probably include definition of the scope and limits of the discovered truth or invented technology.

We also need to retune our emphasis as we transit from iron – silicon and pure semiconductor age to carbon based age which promises access to far more complex structures of matter than what we have been used to. This will change the rate at which we expand our base of science and technology. In figure 6 we have shown the growth to be plateaued but that may well be a short term phase. We may well re-start an exponential phase of development thereafter. The future orientation of future funding of science will have to worry about these issues and future institutes that emphasize applied research will have to focus on these aspects of the coming phase of science.

3. Conclusion

We have analysed the evolution of human studies of nature from early ad hoc approach to formal scientific methods. The latter can be of three kinds, pragmatic, axiomatic and logic based approach. We discussed the relevance and important contribution of each system. We then show that the pragmatic and axiomatic approaches, though highly successful in their times, are at the limit of their ability to explore nature and the coming generation of scientific studies will more in the form of logic based approach where formal proofs from first principle will no longer be possible and simulation and experimentation will be the primary methods of building up our knowledge base about the working of nature.

| No | Approach | Period of dominance | Characteristics | Approach to observation of smoke on the mountain | Major Achievements |
|---|---|---|---|---|---|
| | **Table 1: Different approaches to studies of nature** | | | | |
| 1 | Ad hoc Approach | 2 millennium BC to 5000 BC | Makes working objects based on perceived need | There is smoke on the mountain – avoid the region | Early technologies from stone tools to travel. |
| 2 | Religious Approach | 2000 BC to 1000 AD | Humans are taught required skills by divine intervention when humans are ready for it. | There is a divine smoke on the mountain – worship it | Stabilisation of society. |
| 3 | Pragmatic Approach | 3000 BC to 1600 AD and continuing to date but at a lower scale | Nature works in logical and consistent ways that can be analysed. But any such knowledge is topical and good only for the situation in which it is applied. | There is smoke so there must be fire on the mountain | Clarity and mathematical precision in prediction of seasons to all aspects of human existence |
| 4 | Axiomatic Approach | 1600 AD onwards | Nature's working is consistent and universal and nature obeys all its rules under all conditions and has no exceptions. | The smoke on the mountain implies that: 1) There is dry inflammable material on the mountain. 2) There is a source of heat that heated this material to the temperature where is caught fire. | Development of new technologies, simplified description of nature. |
| 5 | Logic-based Approach | 500 AD onwards but less prominent than Axiomatic approach | Working of nature is logical and consistent that extends to common rules which work well. However, there is no admission of generalised universal laws. | There is smoke so there is fire, implying that there is inflammable material on the mountain. | Providing intellectual explanation for the working of the laws of nature. |

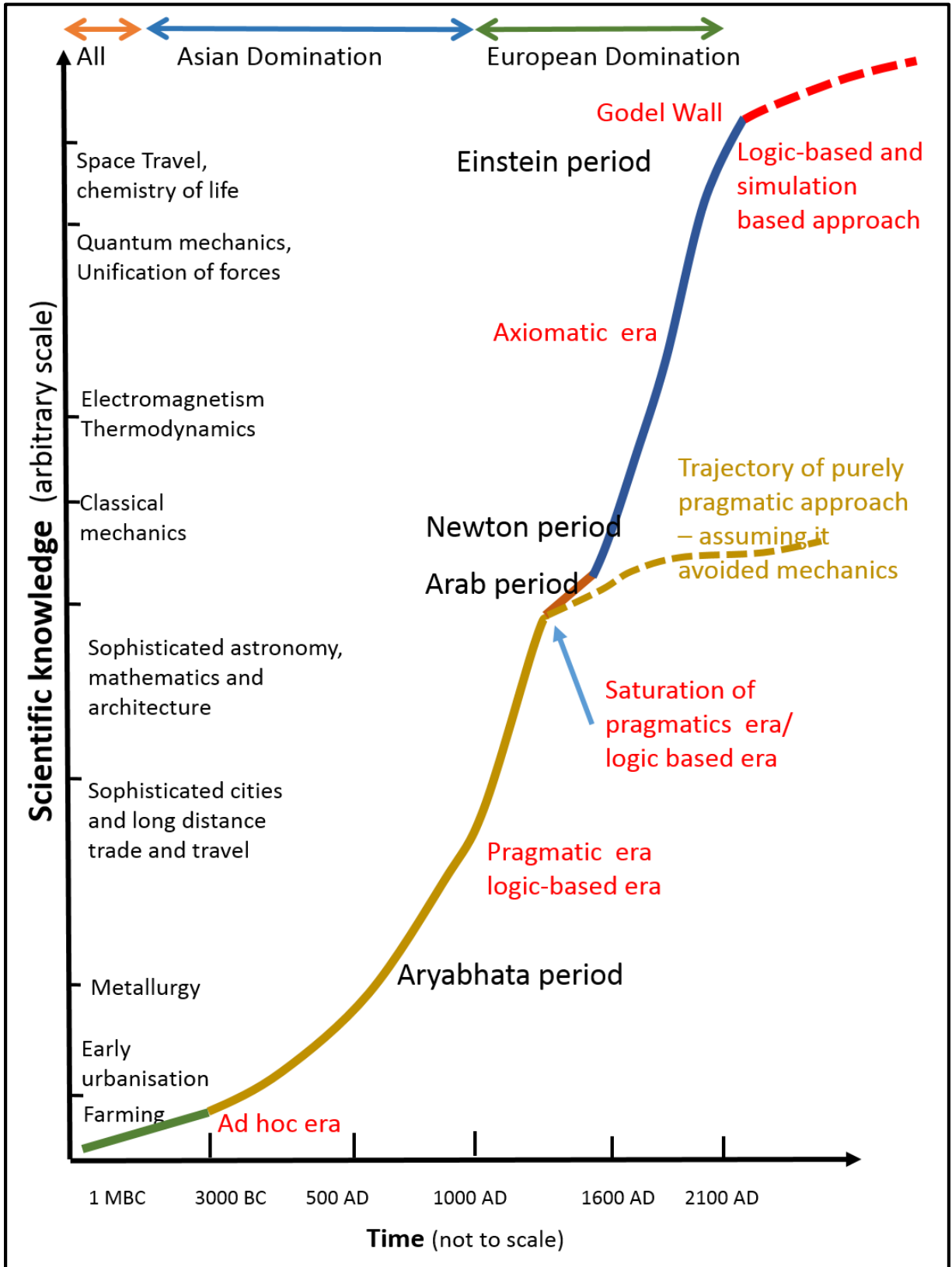

*Figure 2: Sketch of the most dominant approach to science over human civilisation*

Acknowledgement

The author wishes to thank Sir Arnold Wolfendale for his continuing discussions and incisive questions which have helped clarify several aspects of the perspective presented here. He also wishes to thank Prof. Deepak Mathur whose careful reading of the manuscript and valuable inputs made the manuscript more exhaustive. We also wish to acknowledge the contribution of Mr. V Nandagopal in critically reading the manuscript. The author also wishes to thank Prof. Roddam Narasimha for several illuminating discussions during the course of the study. The author also wishes to thank Nisha Yadav for their contribution in evolving the manuscript.

Evolution of science II: Insights into working of Nature